# Breakdown characteristics of $SiN_x$ with different stoichiometries for resistive memories*

A.E. Mavropoulis, I. Kanellopoulos, G. Pissanos, G. Samara, N. Vasileiadis, E. Stavroulakis, P. Normand, G. Ch. Sirakoulis, P. Dimitrakis

***Abstract*— The breakdown characteristics of $SiN_x$ layers with different stoichiometries are explored. The stoichiometry of $SiN_x$ is modified by changing the gas flow rates during the LPCVD deposition. These layers are suitable for RRAM cells.**

## I. Introduction

Silicon nitride ($Si_3N_4$) is commonly used for charge-trapping (CT) nonvolatile memories for many years [1], [2] in Flash SONOS [3] and Vertical CT-NVM technologies [4]. Defects are present in amorphous $SiN_x$ layers, which are caused by the deficiency of nitrogen due to the thermodynamics and conditions of the deposition methods. These intrinsic defects have different configurations, i.e., nitrogen vacancies, Si-dangling bonds etc. [5]. Resistive memories (RRAMs) with $SiN_x$ as switching material have been demonstrated [6], [7]. The effect of the nitride layer's stoichiometry x=[N]/[Si] as well as the nature of electrode material have been investigated in [6], [7] and [8], [9] respectively. $SiN_x$ RRAMs are promising candidates for memory cell scaling [10], neuromorphic, in-memory and edge computing [11], [12], [13] as well as security applications [14], [15] realizing true-random number generators and physical unclonable functions. It is possible to program the $SiN_x$ RRAMs at different low resistance states (LRS), which are well distinguished from the high resistance state (HRS).

The conductive filament in $SiN_x$ RRAMs is created by nitrogen vacancies [16], so it is possible to assume that changing the stoichiometry of the devices will affect the number of available vacancies that form the filament. This change is achieved in this work by modifying the gas flow rates during LPCVD deposition.

## II. Experimental

By utilizing LPCVD with dichlorosilane (DCS) and ammonia ($NH_3$), $SiN_x$ layers were deposited on three heavily Phosphorus-doped SOI pieces (100nm Si and 200nm BOX) [17]. The flow rates of the precursor gases were modified according to Table 1. This led to the fabrication of a stoichiometric reference sample, a nitrogen rich and a silicon rich when compared to the stoichiometric. The deposition temperature and time remained constant, 800°C and 100s respectively. The top electrode (TE) is created by depositing 30nm Cu and 30nm Pt on top via sputtering, to prevent oxidation. The bottom electrode (BE) consists of 100nm Al that was thermally evaporated. A schematic of the fabricated devices, as well as typical I-V curves are presented in Figure 1.

A stoichiometric $SiN_x$ is deposited on bulk Si substrate for comparison.

**Table 1** Dependence of dielectric properties of $SiN_x$ films on the flow rate of precursor gases

| Sample | DCS (sccm) | $NH_3$ (sccm) | Thickness (nm) | Dielectric const. ε @0.1MHz | σ' (μS) @0.1MHz |
|---|---|---|---|---|---|
| **Stoich.** | 20 | 60 | 8.8 | 6.79 | 1.24 |
| **N-rich** | 20 | 20 | 7.2 | 5.83 | 1.15 |
| **Si-rich** | 60 | 60 | 11.5 | 7.13 | 1.15 |

*Research supported by Hellenic Foundation of Research & Innovation (HFRI) under project LIMA-CHIP (Project no. 2748)

A.E. Mavropoulis, I. Kanellopoulos, G. Pissanos, G. Samara, and P. Normand are with the Institute of Nanoscience and Nanotechnology (INN) in NCSR "Demokritos", Ag. Paraskevi 15341, Greece (a.mavropoulis@inn.demokritos.gr, dkan99@hotmail.gr, g.pissanos@inn.demokritos.gr, g.samara@inn.demokritos.gr, p.normand@inn.demokritos.gr)

N. Vasileiadis and P. Dimitrakis were with INN-NCSR "Demokritos" and currently are with ISD S.A. and the Institute of Quantum Computing and Quantum Technology (n.vasiliadis@inn.demokritos.gr, p.dimitrakis@qi.demokritos.gr) respectively.

A.E. Mavropoulis, E. Stavroulakis and G. Ch. Sirakoulis are also with the Department of Electrical and Computer Engineering, Democritus University of Thrace, Xanthi 67100, Greece (emmastav5@ee.duth.gr, gsirak@ee.duth.gr).

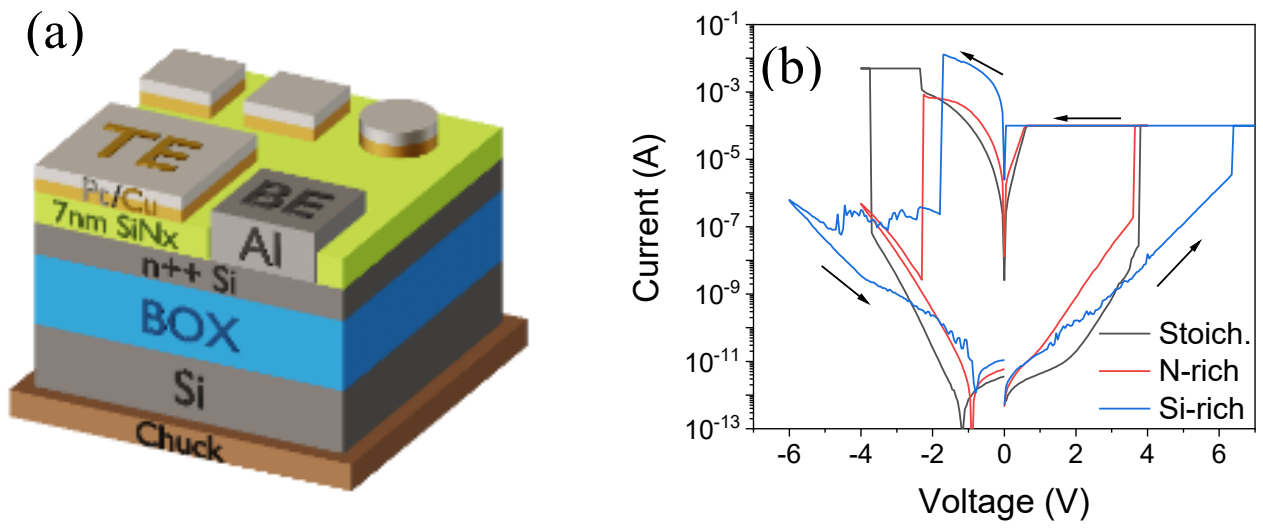

**Figure 1** (a) Device structure description and (b) comparative plots of I-V sweeps for the tested MIS RRAMs with silicon nitride films.

## III. Results and Discussion

A series of electrical characterization techniques, as well as ellipsometry will be used to understand the characteristics of the fabricated devices.

### *A. $SiN_x$ characterization*

Initially, the layer thickness is measured using ellipsometry and the results are shown in Table 1. Obviously, the greater the flow of DCS the thicker the film is, for a constant $NH_3$ flow (60 sccm). Also, the lower the $NH_3$ flow, the thinner the deposited film is, for a constant DCS flow (20 sccm). The index of refraction for N-rich is lower than the stoichiometric and for the Si-rich it's higher. According to [18] higher concentrations of Si lead to a higher index of refraction. Impedance spectroscopy measurements revealed the dielectric constant, $\varepsilon$, and the conductivity of the films (Table 1).

Fourier Transform Infrared (FTIR) spectroscopy plays a key role in identifying structural features, bonding configurations, and impurities in $Si_3N_4$, particularly in thin films. In Figure 2, the FTIR transmission spectra for the fabricated films are presented. As expected, the hydrogen and/or humidity content is very low according to the intensity of Si-H (ca. 2280 $cm^{-1}$) and N-O, O-H absorption bands (ca. 3250 $cm^{-1}$). The presence of the low intensity Si-O-Si vibration peak (ca. 1150 $cm^{-1}$) denotes the presence of a very thin interfacial layer. XTEM studies cannot reveal this layer, suggesting that it is restricted to a few atomic layers. The amorphous nature of the deposited films is defined by the strong Si-N absorption band (ca. 840 $cm^{-1}$). Finally, a very sharp peak was detected at 610 $cm^{-1}$ corresponding to vibrations of Si-C bonds, originating from the carbon contamination in room ambience.

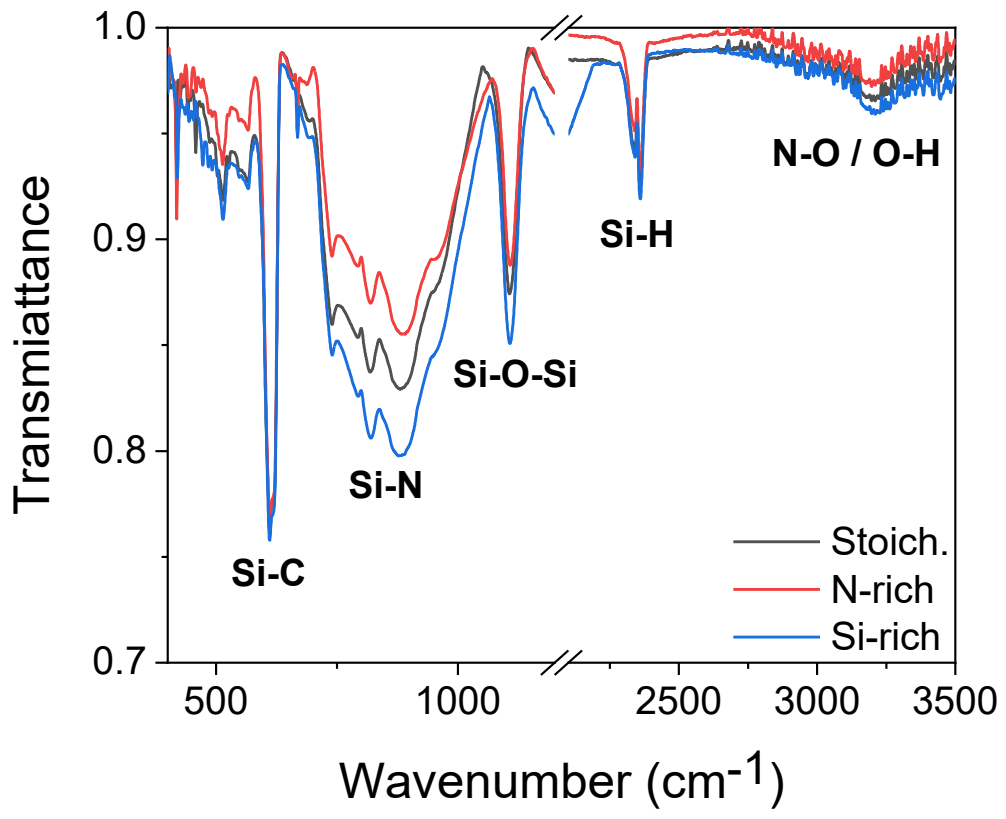

**Figure 2** FTIR transmission spectra obtained for all tested samples.

### *B. Constant voltage stress (CVS)*

All the devices were stressed using a constant voltage and the current was measured every 100ms.

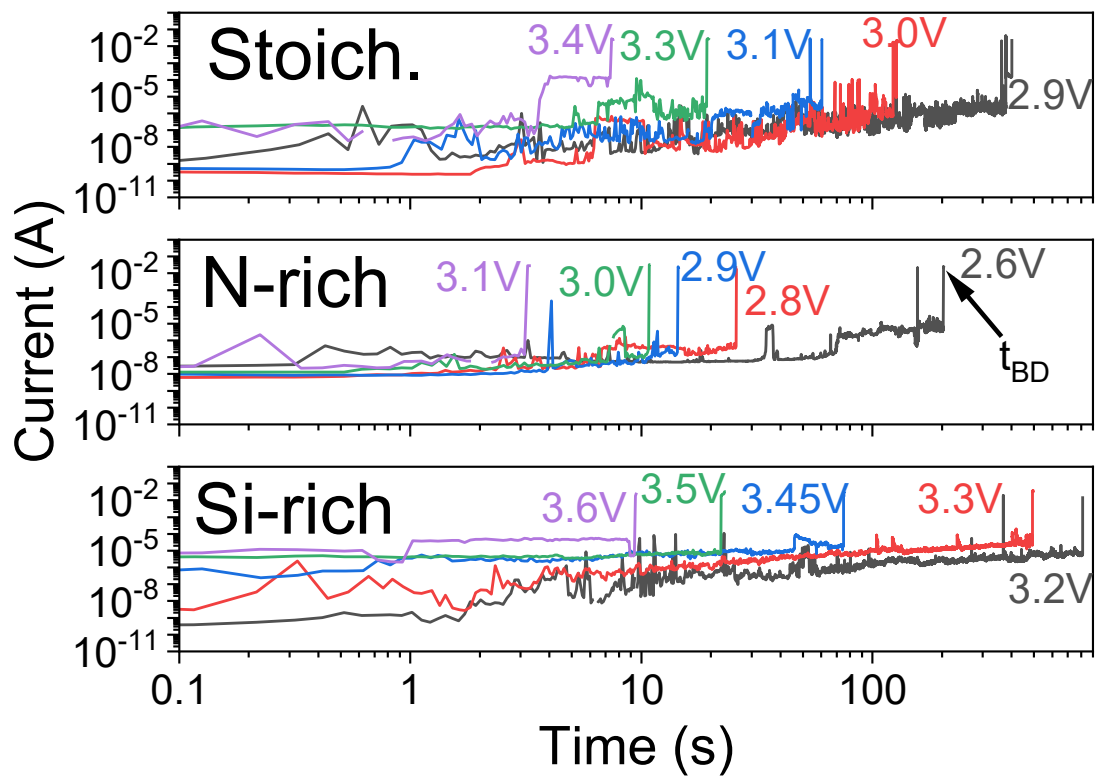

**Figure 3** Constant voltage stress breakdown.

The current evolution under CVS breakdown tests in the examined LPCVD $SiN_x$ films is demonstrated in Figure 3. In Figure 4, the $t_{BD}$ dependence with $1/E_{Stress}$, $E_{Stress}$ and $(E_{Stress})^{1/2}$ is demonstrated. Due to the different thickness of each $SiN_x$, it is important to show the dependance with $E_{Stress}$ instead of $V_{Stress}$ for a better comparison. Data in Fig.4(a) exhibit slightly better linear fitting correlation factor compared to the others. In case of intrinsic breakdown, the $1/E_{Stress}$ model indicates a Fowler-Nordheim tunneling mechanism and $(E_{Stress})^{1/2}$ refers to the Poole-Frenkel mechanism [19]. In case of defect related breakdown, according to [20]

$$t_{BD} = Ae^{B(d-\Delta d)/V_{Stress}} = Ae^{B(1-\frac{\Delta d}{d})/E_{Stress}} \qquad (1)$$

where $V_{Stress}$ ($E_{Stress}=V_{Stress}/d$) is the stress voltage (field), d is the thickness of the dielectric, Δd is the effective dielectric thinning due to the defects formed inside the $SiN_x$ and B is a critical electric field. The slope of linear least-square fitting for data in Figure 4(a) is called the electrical field acceleration factor and is equal to

$$\frac{dln(t_{BD})}{d(1/E_{Stress})} = B(1-\frac{\Delta d}{d}) \qquad (2)$$

This slope increases as the $SiN_x$ becomes richer in Si. The higher the $E_{Stress}$, the shorter $t_{BD}$ is. However, the voltages to breakdown ($V_{BD}$) are overall higher for Si-rich $SiN_x$ compared to the other samples which is mainly attributed to the higher thickness of this film. If the electric field is calculated, the stoichiometric $SiN_x$ requires about 3.5MV/cm to breakdown, the N-rich 4.3MV/cm and the Si-rich 2.7MV/cm. The existence of more traps in the Si-rich sample can explain the lower fields required to break down. Nevertheless, the measured MIS samples exhibited DC I-V characteristics which are fitted to P-F as well as to Modified Space Charge Limited Current (MSCLC) models [21]. In both models the current depends on the *exp(V^0.5^)*.

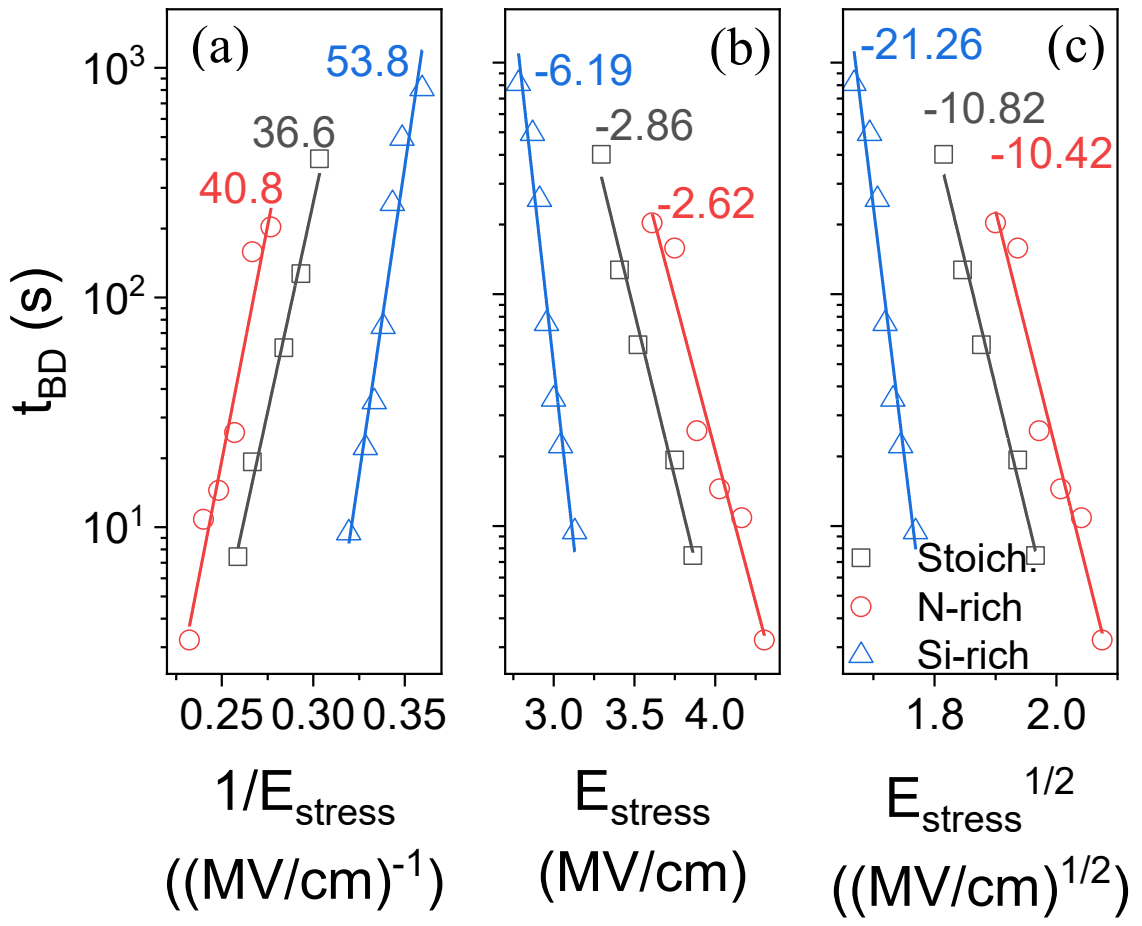

**Figure 4** Time of breakdown, $t_{BD}$, versus (a) $1/E_{Stress}$, (b) $E_{Stress}$ and (c) $(E_{Stress})^{1/2}$. The slope of each line is denoted in the graph.

The charge required until dielectric breakdown occurs, $Q_{BD}$, is calculated by integrating the current transient under CVS of Figure 3 for each stress voltage. A linear dependence between the charge and the electric field is observed in Figure 5. As the $SiN_x$ gets more Si-rich, less charge is needed to cause a breakdown when the electric field is increased. In addition, the presence of increased concentration of Si dangling bonds leads to higher trap-to-trap electronic current leading to shorter $t_{BD}$.

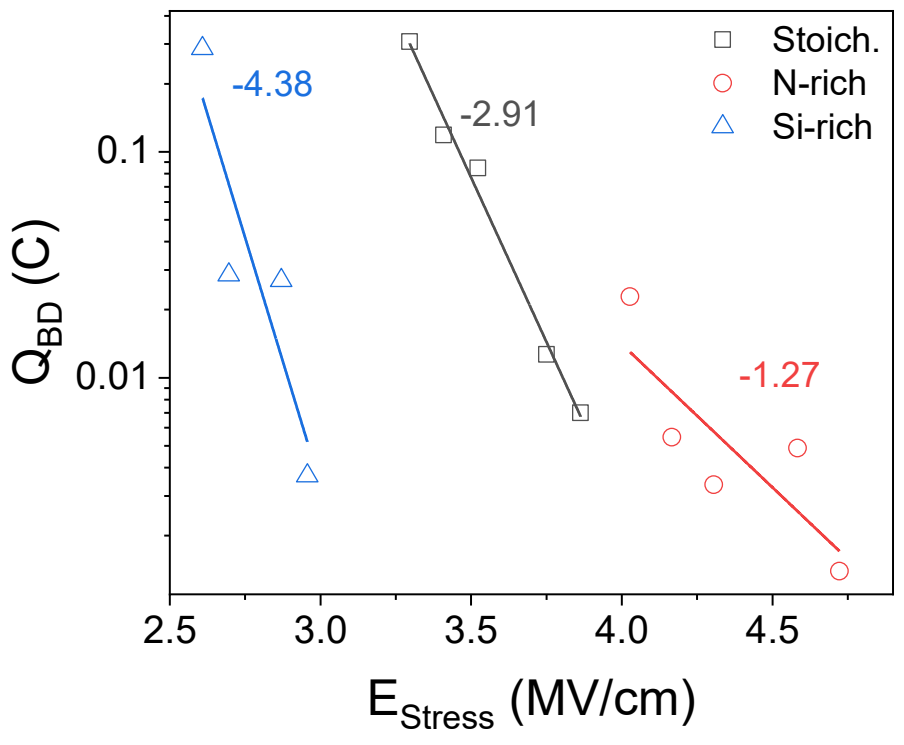

**Figure 5** Charge required to breakdown for each $E_{Stress}$.

By looking closely at the current transient before dielectric (hard) breakdown, RTN signals can be observed, as shown in Figure 6(a). According to the RTN theory [22], [23]

$$\frac{\tau_c}{\tau_e} = exp\left(-\frac{qx_T}{kT}\frac{V}{d}\right) \tag{3}$$

where d is the thickness of silicon nitride films. By plotting the logarithm of the ratio between the capture and emission time, $t_c/t_e$, versus the $V_{Stress}$, as shown in Figure 6(b), the depth of the traps responsible for the observed RTN is calculated. In addition, a positive slope is attributed to traps interaction with the TE, while with a negative to traps interaction with the BE. The traps interacting with the TE are electron induced and the traps interacting with the BE are induced by nitrogen vacancies [22]. For the stoichiometric and the Si-rich the trap depth is 5.6Å and they interact with the TE. For the N-rich the trap depth is 2.9Å and they interact with the BE. These traps are very close to the interface between the $SiN_x$ and the electrodes.

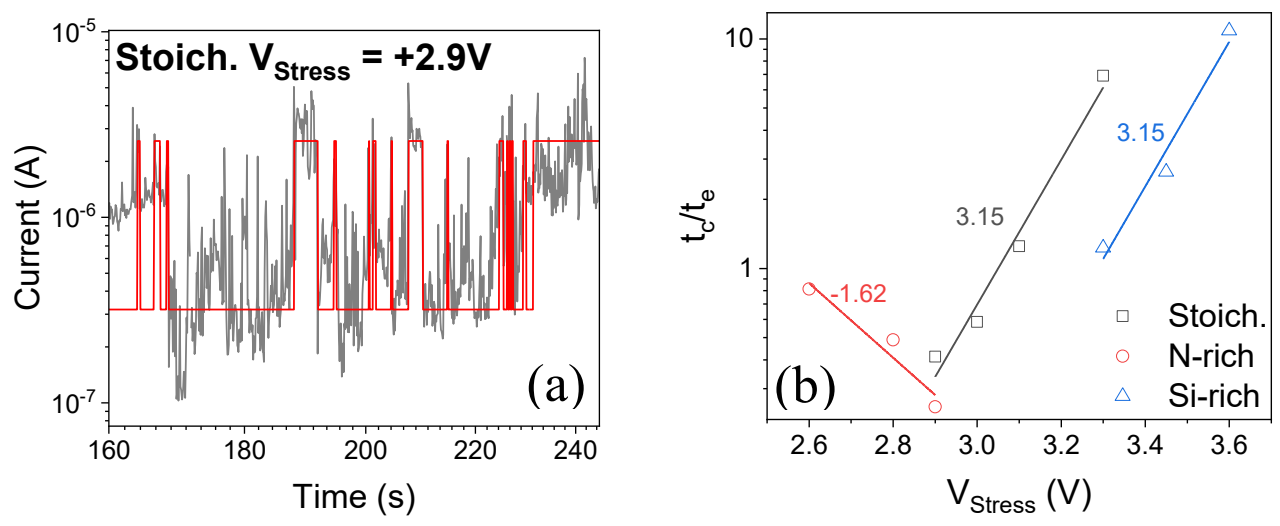

**Figure 6** (a) Indicative RTN signal observed during CVS and (b) plot of the ratio $t_c/t_e$ versus $V_{Stress}$.

Furthermore, a systematic study of the current transients during CVS reveals that before hard breakdown occurs, the current increase and the signal becomes very noisy. A typical example is shown in Figure 7(a). This is due to the strong accumulation of traps around the percolation current path; the conductivity path increases further leading to hard breakdown. The slopes of the current transient plots under CVS before breakdown are presented in Figure 7(b). The steeper the slope, the higher the trap generation rate is.

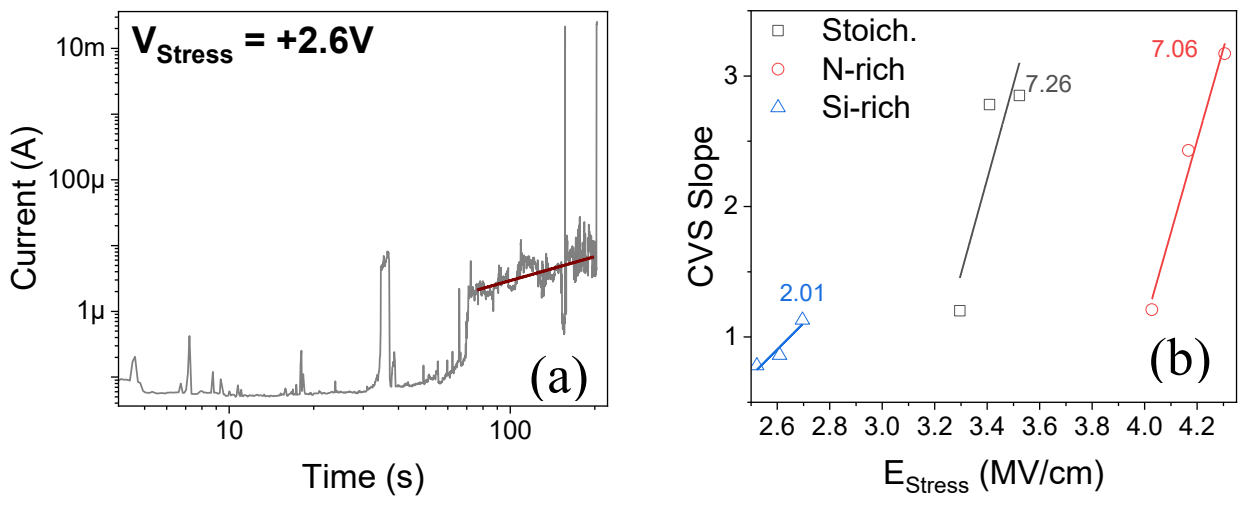

**Figure 7** (a) Typical slope of current transient plot for the N-rich under +2.9V CVS and (b) the transient slope versus $E_{Stress}$.

## C. *SOI vs Bulk Si substrate*

The same methodology as before is followed to compare the breakdown characteristics of the stoichiometric $SiN_x$ on SOI and bulk Si substrates.

Higher voltages are required to breakdown the $SiN_x$ films on SOI substrates. This can be explained by the higher series resistance of the thin 100nm Si layer of the SOI substrate, compared to the bulk Si. Moreover, the bulk Si devices breakdown more quickly by increasing the $V_{Stress}$, as confirmed by the steeper slopes in Figure 8.

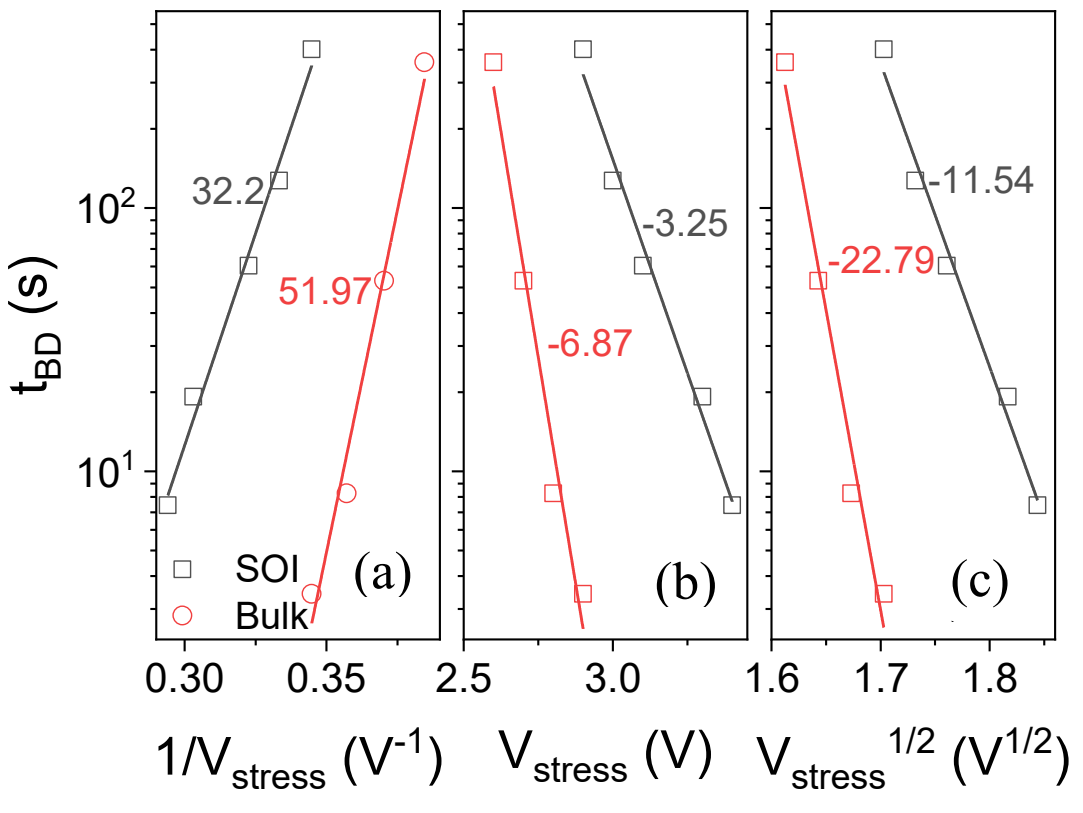

**Figure 8** Time of breakdown, $t_{BD}$, versus (a) $1/V_{Stress}$, (b) $V_{Stress}$ and (c) $(V_{Stress})^{1/2}$ for SOI and bulk Si substrates.

Also, as shown in Figure 9, less charge is required for $SiN_x$ films on bulk Si to break down as $V_{Stress}$ increases.

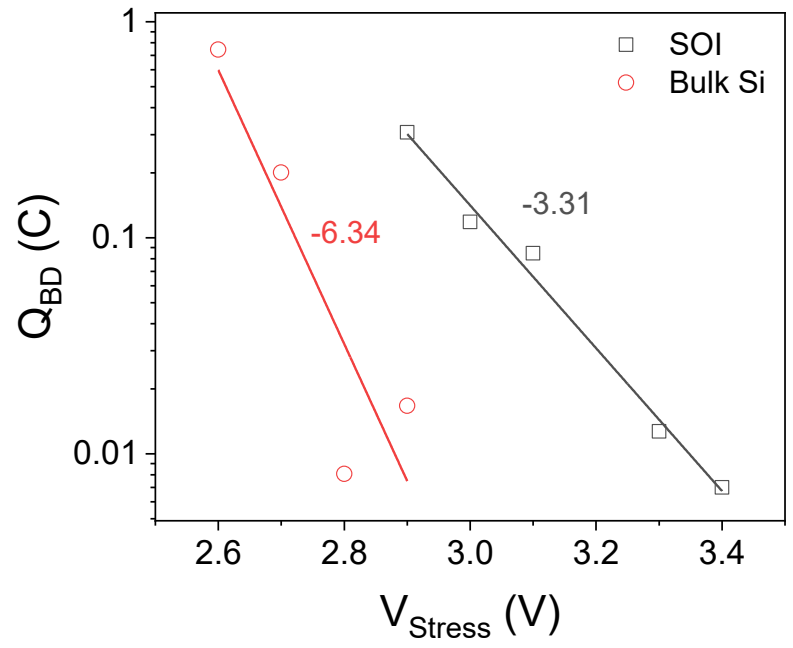

**Figure 9** Charge to breakdown ($Q_{BD}$) as a function of $V_{Stress}$ on SOI and bulk Si substrates.

During breakdown RTN is observed for both substrates. The slope of *log(t<sub>c</sub>/t<sub>e</sub>)* versus the breakdown voltage is positive for both SOI and bulk Si devices, as presented in Figure 10(a), suggesting trap interaction with the TE. Nevertheless, the slope for films on bulk Si is smaller, with the trap depth being calculated at 2.7Å, compared to 5.6Å for the SOI.

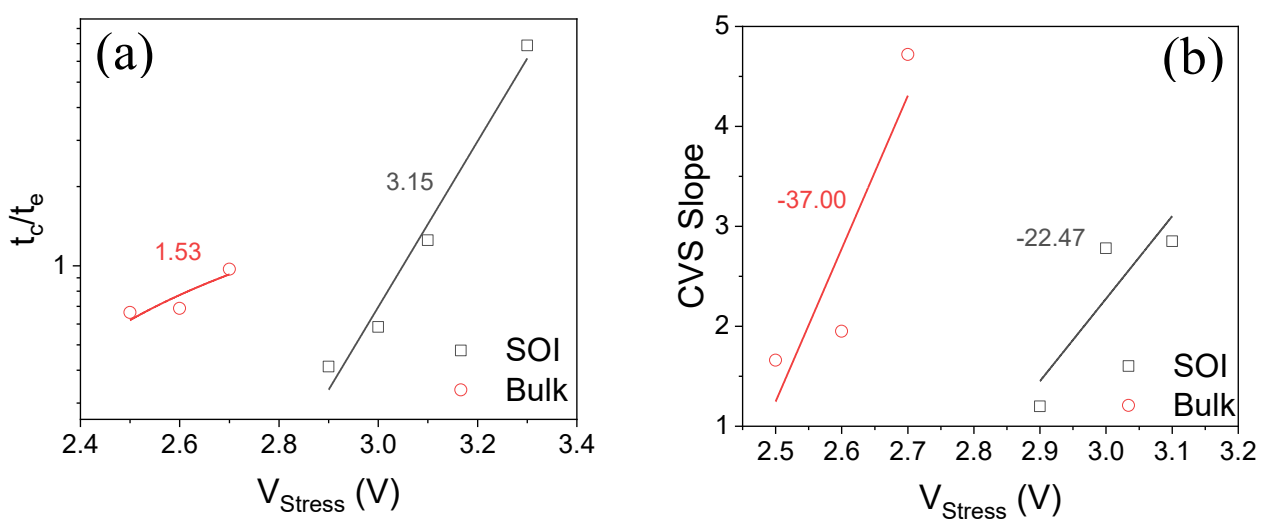

**Figure 10** a) Plot of $\log(t_c/t_e) - V_{Stress}$ and b) slope of linear regression line versus $V_{Stress}$ for films on SOI and bulk Si substrates.

The slope of the current transient plots under CVS in Figure 10(b) is steeper for the bulk Si devices, which means the devices break down more abruptly. This could be the result of higher intensity current flowing through the dielectric, because of the lower series resistance of the substrate compared to the SOI.

## IV. CONCLUSION

$SiN_x$ RRAMs with different stoichiometries (one stoichiometric, one richer in nitrogen and one richer in silicon compared to the stoichiometric) were successfully fabricated. These devices were stressed using constant voltage and RTN was observed during the breakdown measurement. By analyzing the RTN it was found that the traps formed in the dielectric are very close to the interfaces: for the stoichiometric and Si-rich RRAMs they interact with the TE, while for the N-rich with the BE. As the $SiN_x$ gets richer in Si, lower electric fields are required to cause a breakdown, due to the existence of more traps and the time to breakdown decreases more easily as $E_{Stress}$ increases. In addition, the higher series resistance of the SOI substate slightly improves the breakdown characteristics when compared to devices on bulk Si.

## ACKNOWLEDGMENT

This work was financially supported by the research project “LIMA-chip” (Proj.No. 2748) of the Hellenic Foundation of Research and Innovation (HFRI).

## REFERENCES

[1] P. Dimitrakis, “Introduction to NVM Devices,” in *Charge-Trapping Non-Volatile Memories*, Cham: Springer International Publishing, 2015, pp. 1–36. doi: 10.1007/978-3-319-15290-5_1.

[2] J. Brewer and M. Gill, *Nonvolatile Memory Technologies with Emphasis on Flash*. Wiley, 2007. doi: 10.1002/9780470181355.

[3] K. Ramkumar, “Charge trapping NVMs with metal oxides in the memory stack,” in *Metal Oxides for Non-volatile Memory*, Elsevier, 2022, pp. 79–107. doi: 10.1016/B978-0-12-814629-3.00003-9.

[4] H.-T. Lue, “3D NAND Flash Architectures,” in *Charge-Trapping Non-Volatile Memories*, Cham: Springer International Publishing, 2015, pp. 103–163. doi: 10.1007/978-3-319-15290-5_4.

[5] C. Di Valentin, G. Palma, and G. Pacchioni, “Ab Initio Study of Transition Levels for Intrinsic Defects in Silicon Nitride,” *The Journal of Physical Chemistry C*, vol. 115, no. 2, pp. 561–569, Jan. 2011, doi: 10.1021/jp106756f.

[6] S. Kim, Y.-F. Chang, M.-H. Kim, and B.-G. Park, “Improved resistive switching characteristics in Ni/SiNx/p++-Si devices by tuning x,” *Appl Phys Lett*, vol. 111, no. 3, p. 033509, Jul. 2017, doi: 10.1063/1.4985268.

[7] S. Kim, S. Cho, K.-C. Ryoo, and B.-G. Park, “Effects of conducting defects on resistive switching characteristics of SiN *x* -based resistive random-access memory with MIS structure,” *Journal of Vacuum Science & Technology B, Nanotechnology and Microelectronics: Materials, Processing, Measurement, and Phenomena*, vol. 33, no. 6, Nov. 2015, doi: 10.1116/1.4931946.

[8] S. KIM, S. JUNG, M.-H. KIM, S. CHO, and B.-G. PARK, “Resistive Switching Characteristics of Silicon Nitride-Based RRAM Depending on Top Electrode Metals,” *IEICE Transactions on Electronics*, vol. E98.C, no. 5, pp. 429–433, 2015, doi: 10.1587/transele.E98.C.429.

[9] S. M. Hong, H.-D. Kim, H.-M. An, and T. G. Kim, “Effect of Work Function Difference Between Top and Bottom Electrodes on the Resistive Switching Properties of SiN Films,” *IEEE Electron Device Letters*, vol. 34, no. 9, pp. 1181–1183, Sep. 2013, doi: 10.1109/LED.2013.2272631.

[10] S. Kim *et al.*, “Scaling Effect on Silicon Nitride Memristor with Highly Doped Si Substrate,” *Small*, vol. 14, no. 19, May 2018, doi: 10.1002/smll.201704062.

[11] N. Vasileiadis, V. Ntinas, G. Ch. Sirakoulis, and P. Dimitrakis, “In-Memory-Computing Realization with a Photodiode/Memristor Based Vision Sensor,” *Materials*, vol. 14, no. 18, p. 5223, Sep. 2021, doi: 10.3390/ma14185223.

[12] S. Kim, H. Kim, S. Hwang, M.-H. Kim, Y.-F. Chang, and B.-G. Park, “Analog Synaptic Behavior of a Silicon Nitride Memristor,” *ACS Appl Mater Interfaces*, vol. 9, no. 46, pp. 40420–40427, Nov. 2017, doi: 10.1021/acsami.7b11191.

[13] N. Vasileiadis *et al.*, “A New 1P1R Image Sensor with In-Memory Computing Properties Based on Silicon Nitride Devices,” in *2021 IEEE International Symposium on Circuits and Systems (ISCAS)*, IEEE, May 2021, pp. 1–5. doi: 10.1109/ISCAS51556.2021.9401586.

[14] N. Vasileiadis, P. Dimitrakis, V. Ntinas, and G. Ch. Sirakoulis, “True Random Number Generator Based on Multi-State Silicon Nitride Memristor Entropy Sources Combination,” in *2021 International Conference on Electronics, Information, and Communication (ICEIC)*, IEEE, Jan. 2021, pp. 1–4. doi: 10.1109/ICEIC51217.2021.9369817.

[15] R. Carboni and D. Ielmini, “Stochastic Memory Devices for Security and Computing,” *Adv Electron Mater*, vol. 5, no. 9, Sep. 2019, doi: 10.1002/aelm.201900198.

[16] A. Mavropoulis *et al.*, “Silicon nitride resistance switching MIS cells doped with silicon atoms,” *Solid State Electron*, vol. 213, p. 108851, Mar. 2024, doi: 10.1016/j.sse.2023.108851.

[17] N. Vasileiadis *et al.*, “Understanding the role of defects in Silicon Nitride-based resistive switching memories through oxygen doping,” *IEEE Trans Nanotechnol*, pp. 1–1, 2021, doi: 10.1109/TNANO.2021.3072974.

[18] R. F. Wolffenbuttel, D. Winship, Y. Qin, Y. Gianchandani, D. Bilby, and J. H. Visser, “Optical properties of nitride-rich SiNx and its use in CMOS-compatible near-UV Bragg filter fabrication,” *Optical Materials: X*, vol. 24, p. 100348, Dec. 2024, doi: 10.1016/j.omx.2024.100348.

[19] T. Remmell *et al.*, “Reliability of silicon nitride dielectric-based metal-insulator-metal capacitors,” in *2004 IEEE International Reliability Physics Symposium. Proceedings*, IEEE, pp. 573–574. doi: 10.1109/RELPHY.2004.1315395.

[20] I. C. Chen, J. Lee, and C. Hu, “Accelerated Testing of Silicon Dioxide Wearout,” in *1987 Symposium on VLSI Technology. Digest of Technical Papers*, Karuizawa, Japan, 1987, pp. 23–24.

[21] A. E. Mavropoulis, G. Pissanos, N. Vasileiadis, P. Normand, G. Ch. Sirakoulis, and P. Dimitrakis, “SiNx RRAMs performance with different stoichiometries,” in *EUROSOI-ULIS 2025*, Warsaw, Poland, 2025.

[22] S. S. Chung, “The discovery of a third breakdown: phenomenon, characterization and applications,” *Applied Physics A*, vol. 129, no. 2, p. 124, Feb. 2023, doi: 10.1007/s00339-023-06383-w.

[23] C. M. Chang *et al.*, “The observation of trapping and detrapping effects in high-k gate dielectric MOSFETs by a new gate current Random Telegraph Noise (IG-RTN) approach,” in *2008 IEEE International Electron Devices Meeting*, IEEE, Dec. 2008, pp. 1–4. doi: 10.1109/IEDM.2008.4796815.